# Maintaining a Large Process Model Aligned with a Process Standard: an Industrial Example

Martín Soto, Jürgen Münch

Fraunhofer Institute for Experimental Software Engineering,
Fraunhofer-Platz 1, 67663 Kaiserslautern, Germany
`{soto, muench}`@iese.fraunhofer.de

**Abstract.** An essential characteristic of mature software and system development organizations is the definition and use of explicit process models. For a number of reasons, it can be valuable to produce new process models by tailoring existing process standards (such as the V-Modell XT). Both process models and standards evolve over time in order to integrate improvements or adapt the process models to context changes. An important challenge for a process engineering team is to keep tailored process models aligned over time with the standards originally used to produce them. This article presents an approach that supports the alignment of process standards evolving in parallel to derived process models, using an actual industrial example to illustrate the problems and potential solutions. We present and discuss the results of a quantitative analysis done to determine whether a strongly tailored model can still be aligned with its parent standard and to assess the potential cost of such an alignment. We close the paper with conclusions and outlook.

## 1 Introduction

Documenting its software development processes is a step that every software organization striving to achieve a high level of process maturity must take sooner or later. One problem that many organizations face when first attempting to perform this crucial task is the lack of appropriate expertise: Documenting a complete set of organization-wide development processes is potentially a very large undertaking, and doing it successfully requires highly specialized knowledge that organizations often lack. For these reasons, customizing an existing standard process model can be an excellent option for many organizations, as opposed to documenting their processes "from scratch". A standard process model (e.g., the German V-Modell XT [1]) offers them a solid framework, which can greatly help to guarantee that the resulting process documentation is complete and detailed enough, and that it is structured in such a way that it is useful to process engineers and process performers alike.

Since tailoring is central to process standard adoption, standard models should ideally offer a mechanism for making adaptations in a systematic way, and for keeping those adaptations separated from, but properly linked to, the original standard. Unfortunately, most existing models have not yet reached the point where they can support this type of advanced tailoring out-of-the-box. Therefore, most customization is performed in practice by directly modifying a copy of the original model until it reflects the practices of a given organization. This way, organizations can quickly get up to speed with their own process definition, requiring only access to a standard process model and its corresponding editing tools (which are often distributed together with the model, or are freely available.)

Although very useful in practice, this type of *ad hoc* process model tailoring also introduces some problems, the largest of which is probably long-term maintenance. As soon as tailoring starts, the organization-specific model and the standard model take different paths, and after some time, they will probably diverge significantly. At some point, every organization relying on a customized process model will be confronted with the problem of deciding if it should try to keep it *aligned* with the standard, or if it should rather maintain it as a completely separate entity.

This decision is not easy at all. On the one hand, maintaining the customized model separately implies that, potentially, many corrections and improvements done at the standard level will not be adopted, and also involves the risk that the practices documented for the organization deviate unnecessarily from mainstream accepted practices. On the other hand, keeping the model aligned with the standard implies integrating changes from the standard into the local documentation at regular intervals, a task that, to our knowledge, is not well supported by existing tools and that can be very expensive and unreliable if performed manually.

We believe that this and other similar problems related to process model maintenance can be greatly mitigated by properly managing the evolution of process models. We have devised our *DeltaProcess* [2, 3] approach for process model difference analysis with this goal in mind. The approach makes it possible to efficiently and reliably identify changes in newer versions of a process model with respect to its older versions. It also makes it possible to perform analyses that classify changes in a model (e.g., a process standard) according to their relevance to another model (e.g, a customized model). We expect that by making use of this information, process engineers will be able to save significant effort and produce much more reliable results when trying to align complex process models.

We are currently conducting a study intended to investigate the above hypothesis. In the study, we are trying to help a company to align a process model, customized over a period of about one and a half years, with its corresponding process standard. The rest of this paper uses this case study as an example to illustrate the problems involved in keeping complex process models aligned. The paper is organized as follows: Section 2 describes the process alignment problem and the challenges it presents to process engineers. Section 3 presents a brief description of our *DeltaProcess* approach. Section 4 describes an analysis we performed as part of our ongoing case study to determine the viability of aligning two large process models. Section 5 closes the paper with conclusions and future work.

## 2 Aligning a Customized Process Model With a Standard

In this section, we provide a more detailed description of the problem that occupies us in our case study, namely, aligning a large industrial-grade, customized process model with the standard from which it was originally derived. In order to provide the reader with a complete view of the problem, we describe the process model standard (the German V-Modell XT), the company performing the customization, and the extent and characteristics of their customized model. The section concludes with a discussion of related work, and of why existing approaches are not completely adequate to solve the problem we are dealing with.

## 2.1 The German V-Modell XT

The V-Modell XT [1] is a prescriptive process model intended originally for use in German public institutions, but finding increasing acceptance in the German private sector. Its predecessor, the so-called V-Modell 97, was developed in the 1990s and released originally only in the form of a text document. The V-Modell XT is the result of a recent effort by a publicly-financed consortium of private companies, and government and research institutions to "modernize" the original V-Modell. This effort included converting the original document-based process description into an actual process model with formalized entities and relationships, creating a set of tools to manage instances of the model in this new representation, and improving and extending the actual model contents.

As of this writing, three major versions of the V-Modell XT have been released, namely 1.0 (finished in January 2005 with a minor update in March 2005), 1.1 (finished in July 2005) and 1.2 (finished in January 2006 but released in May 2006.) Further active development by a team of experts from the development consortium is still ongoing. All V-Modell XT releases are freely available and can be downloaded at no cost from the Internet (see [1].)

For editing purposes, instances of the V-Modell are stored as XML files that can be processed using a set of specialized tools (also freely available as an Internet download). The model is structured as a hierarchy of process entities, each having a number of attributes. Entities can be connected to other entities through a variety of relations. Version 1.2 of the V-Modell XT is comprised of about 2100 process entities with over 5000 attributes, and connected by some 4100 entity relations. The paper documentation generated automatically from this model is 620 pages long. Also, the current model schema contains 38 classes and 43 different types of relations. Most of these numbers are only approximate, but should be able to give the reader a general idea of the size and complexity involved.

## 2.2 A Customized Version of the V-Modell XT

We are performing our case study in the context of a medium-sized (about 1200 employees), privately-held company that is an early adopter of the V-Modell XT. Although information technology is not its main business, this company has a software development division with about 70 employees, which is mainly dedicated to the development and maintenance of the company's own information systems. The idea of introducing the V-Modell XT arose in 2005 as part of a software process improvement effort. Since it was judged that the V-Modell XT in its standard form was not adequate for internal use, the company's software process group started a customization effort at the end of 2005, whose first results were seen a year later with the introduction of the model as official guidance for new development projects. The tailored model is based on version 1.1 of the V-Modell, which was the current version at the time the customization effort was started.

The tailored model differs significantly from the standard V-Modell XT. During customization, more than half of the original entities were erased because they were considered irrelevant for the company. The resulting trimmed model was afterwards extended with a number of new entities. Many of the entities preserved from the original model were also adapted, by changing names and descriptions as necessary to fit the local processes and terminology. Despite the extensive changes, the final model still uses the original V-Modell XT metamodel without modification.

As mentioned above, Version 1.2 of the V-Modell XT was released in May 2006, when the company's process customization effort was already quite advanced. As of this writing (March 2007), no attempt has been made to integrate any of the additions and corrections present in version 1.2 into the company's customized model, although members of the software process group have expressed their interest in doing this at least to some extent. This is currently not a high priority because the customization process was finished only recently, but it is acknowledged that there may be corrections and additions in the new V-Modell XT version that could benefit the tailored model.

Due to the size and complexity of the models involved, it is very difficult to manually determine the actual extension of the changes performed on each one of them, and this, in turn, makes it difficult to estimate the effort involved in aligning the tailored model with the standard. As discussed in the following section, determining the extent of the changes and analyzing them to find those that are suitable for incorporation into the tailored model and those that may lead to conflicts has been, until recently, a mainly manual, and thus potentially expensive and unreliable, process.

## 2.3 Difference Identification in the V-Modell

Comparing source code versions and analyzing the resulting differences is a task software developers perform on a daily basis for a variety of purposes, including sharing of changes, review and analysis of changes done by others, and space-efficient storage of multiple versions of a program. Such comparisons can be performed using widely available software, such as the well-known diff utility present in most UNIX-like operating systems, and other similar programs. Diff relies on interpreting files as being composed of text lines (sequences of characters separated by the newline character) and then finding *longest common subsequences* (LCS) of lines by using an efficient algorithm (see [4] for an example). Lines not belonging to a common subsequence are considered to be differences among the compared files.

In most practical cases, entities in a process model are connected in an arbitrary graph structure (the V-Modell XT is a good example of this). Since LCS algorithms can only operate on sequential structures, it is thus impossible to apply them directly to most process models. Nonetheless, the idea of using diff or a similar LCS-based program on process models is still appealing. The reason is that many useful tools, including most source code versioning systems, rely on an LCS algorithm implementation as their only comparison mechanism, and it would be valuable if these tools would work on process models, as opposed to working only on program source code.

For the the team working on the V-Modell XT, for example, it was necessary to introduce a code versioning system to support collaborative work, since members of the team work separately and in parallel on different aspects of the model's contents. In order to do that, each team member changes a separate copy of the model, and later uses the versioning system to *merge* the changes into the main development branch. The merge operation, however, is based on finding a minimal set of changes using diff, and, thus, requires diff to produce somewhat usable results when applied to the V-Modell XML representation. The V-Modell solution to this problem is to format XML files in a special way, carefully controlling the order of elements in the file, and ingenuously introducing line breaks and comment lines into the XML representation. When working with XML files formatted this way, diff is able to recognize simple changes, like added or deleted entities or changed attributes, as separated groups of inserted, deleted, or changed lines.

Although this approach has effectively enabled the use of collaborative versioning tools for the model's development and maintenance, it is not free of problems. First of all, change integration works mostly correctly when integrating *non-conflicting* sets of changes, i.e., sets of changes that affect completely separate areas of the model. If, on the other hand, the change sets happen to touch the same area of the model (e.g., by altering the same attribute in different ways), a conflict is detected and marked. Solving the conflict requires a human being to look into the XML file where the changes have been merged and correct the conflicting lines manually using a text editor. This is a cumbersome process that requires detailed knowledge of the XML representation.

## 3  The *DeltaProcess* Approach

Considering the problems discussed in the previous section, we developed the *DeltaProcess* approach with the following goals in mind:

- Operate on models based on a variety of schemata. New schemata can be supported with relatively little effort.
- Be flexible about the changes that are recognized and how they are displayed.
- Allow for easily specifying change types that are specific to a particular schema or even to a particular application.
- Be tolerant to schema evolution by allowing the comparison of model instances that correspond to different versions of a schema (this sort of comparison requires additional effort, though.)

We claim that our approach is suitable for *difference analysis* as opposed to just difference identification (i.e., simple comparison). First of all, instead of defining a set of interesting change types in advance, we make it possible for the user to specify the types of changes that interest him in a schema-specific way. Additionally, since we use queries to find changes, it is possible for a user to restrict results to relevant areas of a model, according to a variety of criteria. Finally, postprocessing allows for applying specialized comparison and visualization algorithms to the resulting data, making it possible to display changes at a level of abstraction that is adequate for a specific task.

In this section, we provide a brief description of the *DeltaProcess* approach and its implementation *Evolyzer*. Readers interested in the inner workings of the approach are invited to read [2] and [3].

### 3.1 Description of the Approach

In order to compare models, the *DeltaProcess* approach goes through the following steps:

1. Convert the compared models to a normalized triple-based notation.
2. Perform an identity-based comparison of the resulting models, to produce a so-called *comparison model.*
3. Find relevant changes by using queries to search for patterns in the comparison model.
4. Postprocess the resulting change data, in order to refine the results or produce task-specific visualizations.

We explain these steps in some more detail in the following paragraphs.

The first step normalizes the compared models by expressing them as sets of so-called *statements*. Statements make simple assertions about the model entities (e.g., *e1 has type Activity* or *e1 has name "Design"*), or define relations among entities (e.g., *e1 produces product p1)*. Although we could have defined our own notation for the statements, we decided to use the standard RDF notation [5] for this purpose. Beside the standardization benefits, RDF has the formal properties required by our approach.

In general, using a normalized triple notation has a number of advantages with respect to other generic notations like XML:

- It is generally inexpensive and straightforward to convert models to the notation. Since the set of possible assertions is not limited and can be defined separately for every model, models in arbitrary notations can be converted to RDF without losing information.
- Models do not lose their "personality" when moved to the notation. Once converted, model elements are often still easy for human beings to recognize.
- The results of a basic, unique-identifier based comparison can be expressed in the same notation. That is, comparisons are models, too. Additionally, elements remain easy for human beings to identify even inside the comparison.
- Thanks to normalization, a single, simple pattern notation can be used to describe a large number of interesting changes.

In step 2, two or more normalized models (in our case study, we perform many analyses using a three-way comparison) are put together into a single so-called *comparison model*. In this new model, statements are marked to indicate which of the original models they come from. One central aspect of the comparison model is that it is also a valid RDF model. The theoretical device that makes this possible is called *RDF reification,* and is defined formally in the RDF specification [5]. The main purpose of RDF reification is to allow for statements to speak about other statements. This way, it is possible to add assertions about the model statements, telling which one of the original models they belong to.

Changes appear in the comparison model as combinations of related statements that fulfill certain restrictions. For example, the change *a1's name was changed from "Design" to "System Design"* appears in the comparison model as the statement *a1 has name "Design"* marked as belonging only to the older version of the model, and the statement *a1 has name "System Design"* marked as belonging only to the newer version of the model. Since the number of statements in a comparison model is at least as large as the number of statements in the smallest of the compared models (the three-way comparison model used for the case study contains almost 18,000 statements), automated support is necessary to identify such change patterns reliably. For this reason, in step 3, a pattern-based query language is used to formally express interesting change types as queries. By executing the queries, corresponding changes are identified in the comparison model. There is already a standardized notation (SPARQL, see [6]) to express patterns in RDF models. With minimal adaptations, this notation makes it possible to specify interesting types of changes in a generic way. Our *Evolyzer* system (see Section 3.2) provides an efficient implementation of SPARQL that is adequate for this purpose.

The final step involves postprocessing of the change data obtained in step 3 in order to prepare the results for final display. One important purpose of this step is to allow for applying specialized comparison algorithms to particular model elements. For example, changed text descriptions in the V-Modell can be compared using a word-level, LCS-based algorithm to determine which words were changed. We also use this step to generate a variety of textual and graphical representations of change data.

One important limitation of the *DeltaProcess* approach is the fact that it requires that entities have unique identifiers that are consistent in all of the compared model instances. Otherwise, it would be impossible to reliably compare the resulting statements. Although this limitation may appear at first sight to be very onerous, our experience shows that, in practice, most modeling notations actually contain the identifiers, and most modeling tools do a good job of keeping them among versions. The V-Modell is not an exception, since its entities are always given a universal, unique, aleatory identifier at creation time.

### 3.2 Implementation

Our current implementation, *Evolyzer,* (see Fig. 1) was especially designed to work on large software process models, such as the V-Modell and its variants. Nevertheless, since the comparison kernel implements a significant portion of the RDF and SPARQL specifications (with the remaining parts also planned), support for other types of models can be added with relatively small effort.

The current implementation is written completely in the Python programming language, and uses the MySQL database management system to store models. Until now, we have mainly tested it with various process models, including many versions of the V-Modell (both standard releases and customized versions.) Converted to RDF, the latest released version of the V-Modell (1.2) contains over 13.000 statements, which describe over 2000 different entities. A large majority of the interesting comparison queries on models of this size (e.g., those used for producing the results presented in Section 4) run in less than 5 seconds on a modern PC.

### 3.3 Related Approaches

A number of other approaches are concerned with identifying differences in models of some type. [7] and [8] deal with the comparison of UML models representing diverse aspects of software systems. These works are generally oriented towards supporting software development in the context of the Model Driven Architecture. Although the basic comparison algorithms they present could also be applied to this case, the approaches do not seem to support the level of difference analysis we require.

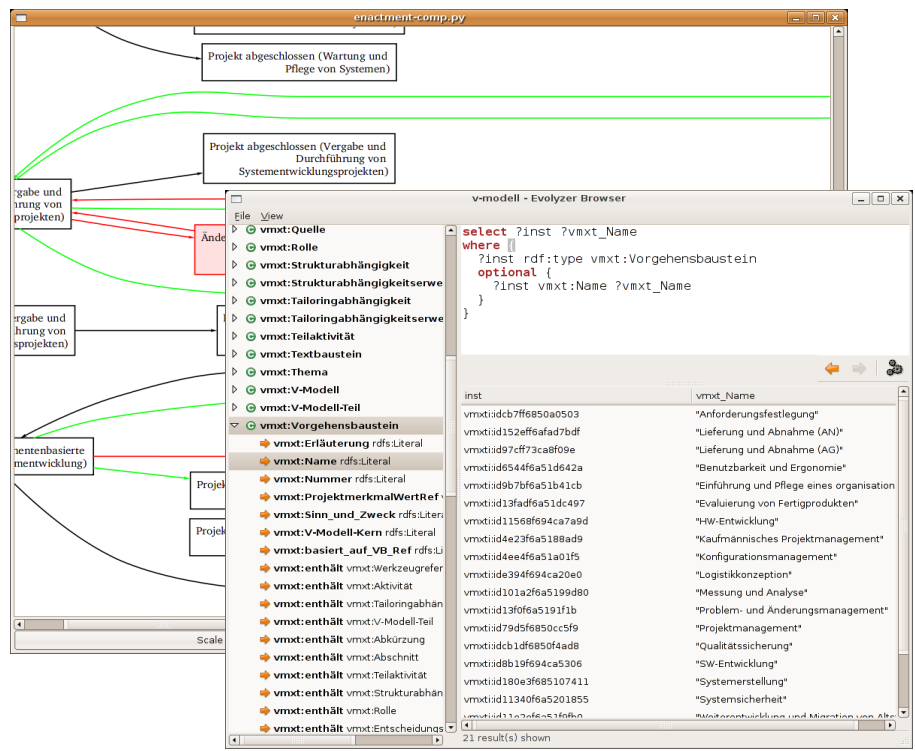

**Fig. 1.** The *Evolyzer* tool working on the V-Modell XT

[9] presents an extensive survey of approaches for software merging, many of which involve a comparison of program versions. Some of the algorithms used for advanced software merging may be applied to the problem of guaranteeing consistent results after a model merge operation, but this is a problem we are not yet trying to solve.

[10] provides an ontology and a set of basic formal definitions related to the comparison of RDF graphs. [11] and [12] describe two systems currently under development that allow for efficiently storing a potentially large number of variants of an RDF model by using a compact representation of the differences between them. These works concentrate on space-efficient storage and transmission of difference sets, but do not go into depth regarding how to use them to support higher-level comparison tasks.

Finally, an extensive base of theoretical work is available from generic graph comparison research (see [13]), an area that is basically concerned with finding isomorphisms (or correspondences that approach isomorphisms according to some metric) between arbitrary graphs whose nodes and edges cannot be directly matched by name. This problem is analogous in many ways to the problem that interests us, but applies to a separate range of practical situations. In our case, we analyze the differences (and, of course, the similarities) between graphs whose nodes can be reliably matched in a computationally inexpensive way (i.e., unique identifiers.)

## 4   An Alignment Viability Analysis

As part of our ongoing case study, we performed an analysis aimed at determining the viability of aligning the company's customized process model with the V-Modell, by incorporating a subset of the changes that occurred in the V-Modell between versions 1.1 and 1.2. In order to perform this assessment, we decided to count the number of entities, entity attribute values, and relations affected by certain types of changes. The purpose of these measurements was to obtain a general impression of the number of separate changes that need to be considered by the process engineers while doing the alignment work.

In order to obtain the values, we defined a change pattern query for every change type, and used the *Evolyzer* tool to execute it and count the results. Although we are only presenting consolidated numbers, the individual changes are available from the tool and could be used by a process engineer as input for the actual alignment task. Regarding effort invested into the analysis, it was performed by one engineer in a single day, with the models having been imported previously into the tool's database.

The table below summarizes our results. The first column numbers the rows for reference, and the second column contains a description of the analyzed change type. The columns labeled "Entities", "Attributes", and "Relations" contain the respective counts of affected model elements. When a change type does not affect a particular type of model element, the corresponding cell remains empty.

| # | Change Type | Entities | Attributes | Relations |
|---|---|---|---|---|
| 1 | Total entities in the V-Modell (1.2) | 2107 | | |
| 2 | Total entities in the tailored model | 1231 | | |
| 3 | Entities present in both models (common entities) | 789 | | |
| 4 | Changed entities in the V-Modell | 536 | 670 | |
| 5 | Common entities changed only by the V-Modell | 96 | 99 | |
| 6 | Common entities containing conflicting attributes | 180 | 210 | |
| 7 | New entities in the V-Modell | 286 | | |
| 8 | New entities in the V-Modell that are contained in preexisting entities | 150 | | |
| 9 | New entities in the V-Modell that are contained in entities still present in the tailored model | 109 | | |
| 10 | Entities deleted from the V-Modell that are still present in the tailored model | 0 | | |
| 11 | New entities in the V-Modell that reference preexisting entities | 170 | | 393 |
| 12 | New entities in the V-Modell that reference entities that are still present in the tailored model. | 100 | | 189 |

| #  | Change Type | Entities | Attributes | Relations |
|----|-------------|----------|------------|-----------|
| 13 | Preexisting entities in the V-Modell that reference new entities | 81 | | 109 |
| 14 | Entities still present in the tailored model that reference new entities in the V-Modell. | 26 | | 41 |
| 15 | New relations between preexisting entities in the V-Modell | | | 67 |
| 16 | New relations in the V-Modell between entities that are also present in the tailored model | | | 7 |
| 17 | Deleted relations (between preexisting entities) in the V-Modell | | | 127 |
| 18 | Relations deleted in the V-Modell between entities still present in the tailored model | | | 1 |
| 19 | Entities in the V-Modell moved to another position in the structure. | 86 | | |
| 20 | Entities still present in both the V-Modell and the tailored model, which were moved by the V-Modell but not by the tailored model | 14 | | |
| 21 | Entities moved to conflicting positions in the structure by the V-Modell and the tailored model | 0 | | |

Rows 1 to 3 present the total entity counts involved. It is clear that the tailoring process deleted a significant portion of the original. Another important observation is that 64% or about two thirds of the entities in the tailored model are still shared with the V-Modell. This portion seems large enough to justify attempting an alignment.

Rows 4 to 6 count the number of changed entities (defined as entities with changed attributes). Lines 5 and 6, in particular, count entities changed by the V-Modell that are still present in the tailored model. The count in 5 (96) corresponds to entities without conflicts, whereas the count in 6 (180) corresponds to entities with conflicts. The sum (276) is the total number of changed entities to consider. Notice that this number is about one half of the total of entities changed by the V-Modell (536). The difference (260) is the number of changed entities that do not have to be considered because they were deleted from the tailored model.

Rows 7-18 try to quantify the size of totally new additions present in the V-Modell. 7 and 8, respectively, count all new entities (286) and new entities contained in preexisting entities. The latter is probably the most relevant count, because the remaining entities are subentities of other new entities, and will probably be considered together with their parents. The subsequent rows try to determine whether it is possible to filter some of these new entities by analyzing their relations to preexisting entities. The resulting values suggest that this is possible, and that a significant number (40 to 50%) can probably be discarded because they have no connections to any of the entities in the tailored model. Line 10, in particular, contains good news: no entity deleted by the V-Modell is still being maintained by the tailored model.

The last three rows (19-21) are an attempt to measure a particular type of structural change, namely, movement of entities in the containment hierarchy. From 86 total changes in the V-Modell, only 14 affect the tailored model, and there are no conflicting changes.

Without historical effort data, it is difficult to produce an exact estimation of the effort involved in performing a model alignment. However, a few conclusions can be extracted from this data. First, integrating the changes done to existing entities (lines 1-3) is probably possible with relatively little effort. Informal observation of the versioning changelogs tells us that many of the changes are small grammar and spelling corrections, but to confirm this, we would need to exactly measure the extent of the changes done to text attributes.

Second, although integrating the new V-Modell elements is likely to take more work, it is also probably viable in a few days time, because the number of entities to consider is relatively small (around 100). Finally, the analysis shows that in this case, the total number of model elements to consider for alignment can be reduced to about half by filtering those elements that were already deleted from the tailored model or that are not connected to elements in the tailored model. This fact alone represents a significant effort saving, which is not achievable with any other method we are aware of.

## 5 Conclusions and Future Work

Organizations trying to document their software processes for the first time may greatly benefit from adopting an existing process standard and customizing it. However, since both process standards and the models derived from them evolve over time, sooner or later they diverge to a point where their lack of alignment becomes problematic. Realigning large process models, however, is a complex problem. Manual alignment is tedious and unreliable, and automated tool support for this task has been insufficient.

Our *DeltaProcess* approach and its *Evolyzer* implementation are a first step to remedy this situation. They provide a framework for identifying changes in process models and for analyzing these changes in order to support particular tasks. The implementation works efficiently on models of the size of the German V-Modell XT.

As the analysis presented in Section 4 shows, our approach can be used effectively to identify relevant changes and filter irrelevant changes when trying to align large process models that were changed independently from each other for an extended period of time. We have not yet started doing the actual alignment as part of our current case study, but expect to be able to attempt it in the following months. A complete experience report will be produced from that effort.

We are also working on extending our tools, which currently concentrate on change analysis, to also support altering the analyzed models. This way, we expect to make it easier for process engineers to work on complex model alignment tasks, by being able to move seamlessly from the change data to the actual model contents.

**Acknowledgments.** This work was supported in part by the German Federal Ministry of Education and Research (V-Bench Project, No.01| SE 11 A).